\documentclass[hidelinks, reprint, amsmath,amssymb, aps]{revtex4-2}

\usepackage{hyperref}
\usepackage{orcidlink}

\usepackage[dvipsnames]{xcolor}

\usepackage{dcolumn}
\usepackage{bm}
\usepackage{physics}
\usetikzlibrary{quantikz2}
\usepackage{amssymb}

\definecolor{pistacchio}{HTML}{C1D37F}
\definecolor{mutedplum}{HTML}{7B3F61}
\definecolor{periwinkle}{HTML}{AFBBF2}

\begin{document}

\preprint{APS/123-QED}

\title{Blind quantum computing with different qudit resource state architectures}

\author{Alena Romanova \orcidlink{0000-0003-0771-8835}}
 \email{alena.romanova@uibk.ac.at}
\author{Wolfgang Dür \orcidlink{0000-0002-0234-7425}}
 \email{wolfgang.duer@uibk.ac.at}
\affiliation{
 Universität Innsbruck, Institut für Theoretische Physik \\ Technikerstraße 21a, 6020 Innsbruck, Austria \\
}

\date{\today}

\begin{abstract}
We discuss how blind quantum computing generalizes to multi-level quantum systems (qudits), which offers advantages compared to the qubit approach. Here, a quantum computing task is delegated to an untrusted server while simultaneously preventing the server from retrieving information about the computation it performs, the input, and the output, enabling secure cloud-based quantum computing. In the standard approach with qubits, measurement-based quantum computing is used: single-qubit measurements on cluster or brickwork states implement the computation, while random rotations of the resource qubits hide the computation from the server. We generalize finite-sized approximately universal gate sets to prime-power-dimensional qudits and show that qudit versions of the cluster and brickwork states enable a similar server-blind execution of quantum algorithms. Furthermore, we compare the overheads of different resource state architectures and discuss which hiding strategies apply to alternative qudit resource states beyond graph states.
\end{abstract}

\maketitle

\section{Introduction}

While quantum computing holds promise to outperform classical computers for various tasks, the experimental availability of a fully fledged quantum computer is limited, and near-term access will likely be through remote servers. Blind quantum computing \cite{Broadbent_2009, introduction-blind-QC} addresses this problem from the user's perspective, enabling a client to delegate a quantum computing task to an untrusted server while maintaining privacy for the computation, the input, and the output.

Experimentally, blind quantum computing with qubits has been demonstrated on photons \cite{ExperimentalDemonstration, Greganti_2016}, on a trapped-ion quantum server with a photonic detection system for the client \cite{Hybdid-BQC-ions-photons}, and on solid-state systems \cite{BlindQCsolidstate}.
In addition, it has been proposed for weak coherent pulses, characterizing also the robustness and security properties of the protocol under possible imperfections \cite{WeakCoherentPulses}. Originally, a single client with the ability to prepare single-qubit states was considered \cite{Broadbent_2009}, but also ideas for multiple clients \cite{BQCmulticlient}, measurement-only clients \cite{BQCmeasurementonlyclient-first, VerificationMeasurementOnlyBQC, BQCmeasurementonlyclient, BQCmeasurementonly-inputverification, IonBQCSimulation}, and more classical clients exist, such as exploiting information flow ambiguity in the cluster state \cite{classical-BQC} or allowing client interaction with multiple non-communicating quantum servers \cite{introduction-blind-QC}. The latter approach was experimentally demonstrated for a factoring problem on photons \cite{PhysRevLett.119.050503}.

The server-blind framework has been extended to continuous-variable systems \cite{CVblindquantumcomputing} and fault-tolerant implementations on the three-dimensional cluster state \cite{BlindTopologicalMBQC, Raussendorf_2007, Raussendorf_3Dcluster_faulttolerant} or the brickwork state \cite{baranes2025designingfaulttolerantblindquantum} have been proposed. Furthermore, the Affleck-Kennedy-Lieb-Tasaki states, which appear as ground states in condensed matter, have been identified as suitable resources \cite{AKLTblindquantumcomputing, MBQC-groundtstates}. In the latter case, the physical systems are qudits, however, the computation is performed in the qubit space.

In turn, we consider blind quantum computing with qudits, so finite-dimensional quantum systems that naturally arise in many physical platforms. Compared to the qubit approach, this can decrease the complexity of quantum circuits \cite{efficientalgorithmswithqudits, quditcircuitcompression, ImprovedCircuitDepthTofolli}, facilitate quantum simulations \cite{LatticeGaugeSimulations, quditsimulation_fermionicsystems}, and enhance fault-tolerant quantum computing \cite{EnhancedFaultTolerantComputing, FaultTolerantQuditSurfaceCode, QuditsIncreasedStability}. Furthermore, utilizing qudits can increase the security of quantum communication \cite{QKDMutuallyUnbiasedBases, QuditQKD_moreSecurity, QuditQKD} and improve the performance of entanglement purification \cite{EntanglementDistillationQudits, EntanglementPurificationQudits} and quantum metrology \cite{QuantumMetrolgyTransmon, ImprovedSensing}.

We show that blind quantum computing can be implemented with qudits, generalizing different qubit resource state architectures, such as the brickwork state, the open-ended cluster state, and the decorated cluster state. For each resource state variant, we demonstrate that both the qubit measurement patterns for gate implementation and the privacy-preserving hiding strategies apply similarly to qudits.
In addition, we compare the required resource overheads and propose a finite-sized approximately universal gate set for prime-power-dimensional qudits, facilitating client-server communication. Hence, our results extend the theoretical foundation of blind quantum computing beyond the qubit regime, opening a pathway towards secure, high-dimensional cloud-based quantum computation.

We start in Sec. \ref{sec:theory} by introducing the required theoretical background, in particular, the mathematical description of qudits, measurement-based quantum computing, and blind quantum computing with qubits. We continue with generalizing blind quantum computing to qudits in Sec. \ref{sec:qudit-blindQC} and conclude in Sec. \ref{sec:conclusion}.

\section{Theoretical background}
\label{sec:theory}

The finite-dimensional state space of qudits can be described in different fashions, and we review the employed formalism in Sec. \ref{subsec:qudits}. Blind quantum computing relies on the measurement-based quantum computing framework, which we introduce in Sec. \ref{subsec:MBQC-qudits}. Afterwards, we discuss in Sec. \ref{subsec:blind-computing} blind quantum computing with qubits, including different strategies for maintaining privacy and how potentially dishonest behaviour of the server is identified.

\subsection{Qudits}
\label{subsec:qudits}

Qudit \cite{QuditComputing, heinrich2021stabiliser, LinearizedStabilizerFormalism} states can be described in different ways. For arbitrary dimensions $d$, we can identify the qudit states with integers in the ring
\begin{equation*}
    \mathbb{Z}_d = \{0, 1, \hdots, d-1 \},
\end{equation*}
where addition and multiplication are performed modulo $d$ \cite{QuditComputing,LinearizedStabilizerFormalism, QuditsArbitraryDimension}. Alternatively, for prime-power dimensions $d = p^m$ with prime $p$ and $m \in \mathbb{N}$ being an integer, one may label the computational basis states via finite-field elements in
\begin{equation*}
    \mathbb{F}_{d} \cong \mathbb{F}_{p}[\xi] / \langle f(\xi) \rangle = \{ a_0 + a_1 \xi + \hdots + a_{m-1} \xi^{m-1} \mid a_i \in \mathbb{Z}_p \}.
\end{equation*}
Here, $\mathbb{F}_p[\xi]$ denotes a polynomial ring in the variable $\xi$ with coefficients from the integer field $\mathbb{Z}_p$ and $f(\xi)$ is an irreducible polynomial (which means that it cannot be factored into non-constant polynomials) of degree $m$ \cite{heinrich2021stabiliser}. In the finite field $\mathbb{F}_{d}$, computations, such as addition or multiplication, are performed modulo the characteristic $p$ and modulo the irreducible polynomial $f(\xi)$.

In the following, we introduce the finite-field description, while the inter-ring version for arbitrary finite dimensions is described in Appendix \ref{app:integer-ring-qudits}. For prime dimensions, these two formalisms coincide.

The finite-field Pauli gates $X(x)$ and $Z(z)$, where $x, z \in \mathbb{F}_d$, act on the qudit basis states according to \cite{heinrich2021stabiliser}
\begin{equation*}
    Z(z) \ket{u} = \chi(z u) \ket{u}, \quad X(x) \ket{u} = \ket{u + x}.
\end{equation*}
Here, $\chi(t) = \omega_p^{\tr(t)}$ with $\omega_p = e^{\frac{2 \pi i}{p}}$ and the finite-field trace $\tr(t)$ is the trace of the multiplication map with $t \in \mathbb{F}_{d}$, transforming qudit basis states into integers via
\begin{equation*}
    \tr(t): \mathbb{F}_{d} \mapsto \mathbb{Z}_p, \quad a \mapsto \tr(a) = \sum_{j=0}^{m-1} a^{p^j}.
\end{equation*}
It holds that
\begin{equation}
\begin{aligned}
    & Z(z) X(x) = \sum_{u \in \mathbb{F}_{d}} \chi(z (u+x)) \ket{u+x} \bra{u}
    \\ & = \chi(z x) \sum_{u \in \mathbb{F}_{d}} \ket{u+x} \bra{u} \chi(z u)
     = \chi(z x) X(x) Z(z).
\end{aligned}
\label{eq:commutation-finitefield-Paulis}
\end{equation}

Finite-field Clifford gates map Pauli gates onto Pauli gates, while preserving the commutation relation in Eq. \eqref{eq:commutation-finitefield-Paulis} and, in addition, being linear in the argument $(z,x)$ of any Pauli $Z(z) X(x)$. The finite-field Clifford group \cite{heinrich2021stabiliser} is generated by $Z(1)$, $ X(1)$, the controlled-$Z$ gate
\begin{equation*}
    CZ = \sum_{(u,v) \in (\mathbb{F}_{d})^2} \chi(uv) \ket{u} \ket{v} \bra{u} \bra{v},
\end{equation*}
the finite-field Hadamard gate
\begin{equation*}
    H = \frac{1}{\sqrt{d}} \sum_{u,v \in \mathbb{F}_{d}} \chi(u v) \ket{u} \bra{v}
\end{equation*}
and the finite-field phase gate $S(\lambda)$. For $p \neq 2$, the phase gate is given by
\begin{equation}
    S(\lambda) = \sum_{u \in \mathbb{F}_{p^m}} \chi(2^{-1} \lambda u^2) \ket{u} \bra{u}
    \label{eq:finitefield-phasegate-odd}
\end{equation}
and for $p=2$, by
\begin{equation}
    S = \sum_{u \in \mathbb{F}_{2^m}} \chi_4(u^2) \ket{u} \bra{u}.
    \label{eq:finitefield-phasegate-even}
\end{equation}
Here, $\chi_4(t) = i^{\tr_4(t)}$ and we evaluate the trace of the multiplication map, $\tr_4(t):\mathbb{GR}(4,m) \mapsto \mathbb{Z}_4$, in the Galois ring $\mathbb{GR}(4,m)$, an extension of the $\mathbb{Z}_4$ ring with extension degree $m$ that has $4^m$ elements.
Supplementing any single-qudit non-Clifford gate renders the Clifford group approximately universal, so that any unitary can be decomposed to arbitrary precision \cite{heinrich2021stabiliser}. If we allow for continuous-parameter diagonal unitaries in addition to the Hadamard gate and the entangling $CZ$ gate, exact universality, meaning that any unitary can be decomposed exactly, holds both for finite-field and integer-ring qudits \cite{Clark_2006,quditMBQC,quditMBQCstabilizerstates}.

Despite the multiplication gate
\begin{equation}
    M(\lambda) = \sum_{u \in \mathbb{F}_{d}} \ket{\lambda u} \bra{u}
    \label{eq:multiplication-gate}
\end{equation}
frequently being mentioned as a Clifford group generator \cite{heinrich2021stabiliser, LinearizedStabilizerFormalism}, it is redundant in a minimal generating set since it can be expressed via the Hadamard and phase gates \cite{Farinholt_2014}.
In particular, we can write the multiplication gate via \cite{Farinholt_2014, quditMBQCstabilizerstates}
\begin{equation}
    M(\lambda) = H S(\lambda) H S(\lambda^{-1}) H S(\lambda).
    \label{eq:multiplication-gate-decomposition}
\end{equation}
In even prime-power dimensions, we replace $S(\lambda)$ with $M(l^{-1}) S M(l)$, where $\lambda = l^2$ (such an $l$ always exists due to the map $l \mapsto l^2$ being a bijection in $\mathbb{F}_{2^m}$).
This decomposition is useful to derive gate identities later on. Furthermore, $H^2 = M(-1)$, so that, in every dimension, $H^4$ is the identity $I_d$. In Appendix \ref{app:clifford-conjugation}, we provide some useful conjugation relations.

The generalized $X$ and $Z$ gates have complex eigenvalues, implying that they are no longer self-adjoint and therefore not observables anymore. However, their respective eigenvectors still form an orthonormal basis of the qudit Hilbert space. The eigenstates of $Z(z)$ are the qudit basis states $\{ \ket{k_Z} \}_{k \in \mathbb{F}_d}$ while the eigenvectors of $X$ are given by
\begin{equation*}
    \ket{k_X} = H \ket{k_Z} = H X(k) \ket{0_Z} = Z(k) H \ket{0_Z} = Z(k) \ket{0_X}.
\end{equation*}
The $Y$ basis is then defined via $\ket{k_Y} \coloneq S(1) \ket{k_X}$.

\subsection{Measurement-based quantum computing}
\label{subsec:MBQC-qudits}

In measurement-based quantum computing \cite{OneWayQC, MBQCClusterStates, MBQCreview}, generalized to qudits \cite{quditMBQC, Clark_2006}, arbitrary gates are deterministically performed via single-qudit measurements on a multipartite entangled resource state. Usually, this is the cluster state \cite{ClusterState} or the brickwork state \cite{Broadbent_2009}.
More generally, the resources are graph states \cite{GraphStates, hein2006entanglementgraphstatesapplications}, where vertices correspond to qudits, initialized in the equal superposition state $\ket{+} \coloneq \ket{0_X}$ and edges to the application of one entangling controlled-phase gate $CZ$.

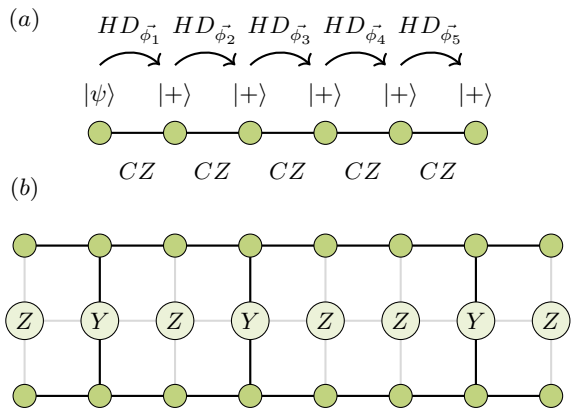
\begin{figure}[t!]
    \centering
    \begin{tikzpicture}

    \node at (-2,1) {$(a)$};

    \draw[->, thick, bend left=45] (-1,0.4) to node[above] {$H D_{\Vec{\phi_1}}$} (-0.2,0.4);
    \draw[->, thick, bend left=45] (0,0.4) to node[above] {$H D_{\Vec{\phi_2}}$} (0.8,0.4);
    \draw[->, thick, bend left=45] (1,0.4) to node[above] {$H D_{\Vec{\phi_3}}$} (1.8,0.4);
    \draw[->, thick, bend left=45] (2,0.4) to node[above] {$H D_{\Vec{\phi_4}}$} (2.8,0.4);
    \draw[->, thick, bend left=45] (3,0.4) to node[above] {$H D_{\Vec{\phi_5}}$} (3.8,0.4);
    
    \draw[thick] (-1,-0.5) -- (4,-0.5);
    \foreach \x in {-1,...,4} {
        \node[circle, fill=pistacchio, draw=black] at (\x,-0.5) {};
    }

    \foreach \x in {-0.5,...,3.5} {
         \node at (\x,-1) {$CZ$};
    }
    
    \node at (-1,0) {$\ket{\psi}$};

     \foreach \x in {0,...,4} {
         \node at (\x,0) {$\ket{+}$};
    }

    \node at (-2,-1.25) {$(b)$};
    
    \foreach \x in {-2,...,5} {
        \draw[thick,opacity=0.15] (\x,-2) -- (\x,-4);
    }
    \foreach \y in {-2,...,-4} {
        \draw[thick,opacity=0.15] (-2,\y) -- (5,\y);
    }

    \draw[thick] (-2,-2) -- (5,-2);
    \draw[thick] (-2,-4) -- (5,-4);

    \draw[thick] (-1,-2) -- (-1,-4);
    \draw[thick] (1,-2) -- (1,-4);
    \draw[thick] (4,-2) -- (4,-4);

    \foreach \x in {-1,1,4} {
        \node[circle, fill=pistacchio!30, text=black, draw=black, minimum size=5mm, inner sep=0pt] at (\x,-3) {$Y$};
    }

    \foreach \x in {-2,0,2,3,5} {
        \node[circle, fill=pistacchio!30, text=black, draw=black, minimum size=5mm, inner sep=0pt] at (\x,-3) {$Z$};
    }
    
    \foreach \x in {-2,...,5} {
        \foreach \y in {-2,-4} {
        \node[circle, fill=pistacchio, draw=black] at (\x,\y) {};
        }
    }
\end{tikzpicture}
    \caption{Measurement-based quantum computing on the qudit cluster state. $(a)$ Single-qudit gates are implemented via local measurements on one-dimensional resource state chains, each processing a logical qudit $\ket{\psi}$. The information flows from left to right, and the state $\ket{\psi}$ is subject to $H D_{\Vec{\phi_k}}$, $k \in \{1,\hdots, 5\}$, in each step. $(b)$ The upper and lower horizontal chains correspond to logical qudits being processed and $Z$ and $Y$ measurements on the physical qudits in between place entangling gates at desired positions while simultaneously deleting the measured qudits from the resource state.
    }
    \label{fig:mbqc-basics}
\end{figure}

The cluster state is structured as a regular two-dimensional lattice. Single-qudit gates are realized on one-dimensional resource state chains with the information, a logical qudit, moving from left to right, as indicated in Fig. \ref{fig:mbqc-basics} $(a)$.

Each single-qudit measurement is performed in the $X$ basis, rotated by a diagonal gate
\begin{equation*}
    D_{\Vec{\phi}} \coloneq \textnormal{diag}(e^{i \phi_1}, \hdots, e^{i \phi_d}) = e^{i \sum_k \phi_k \ket{k_Z} \bra{k_Z}}.
\end{equation*}
If we measure an arbitrary qudit state $\ket{\psi} = \sum_l \alpha_l \ket{l_Z}$, $\sum_l |\alpha_l|^2 = 1$, entangled with $\ket{0_X}$ via $CZ$, in such a rotated $X$ basis with outcome $k \in \mathbb{F}_d$, the quantum information $\ket{\psi}$ effectively moves to the next site while simultaneously being processed with $ H Z(-k) D_{\Vec{\phi}}^\dagger$ since
\begin{equation*}
\begin{aligned}
    & \bra{k_X}_1 D_{\Vec{\phi}}^\dagger \hspace{0.1cm} CZ \left(\ket{\psi}_1 \ket{0_X}_2 \right)
    \\ & = \bra{0_X}_1 Z(-k) D_{\Vec{\phi}}^\dagger \sum_{k} \alpha_l \ket{l}_1 H \ket{l}_2
    \\ & \propto H Z(-k)  D_{\Vec{\phi}}^\dagger \ket{\psi}_2 = X(k) H D_{\Vec{\phi}}^\dagger \ket{\psi}_2.
\end{aligned}
\end{equation*}

Repeated measurements along one-dimensional chains then realize a sequence of $ \{ H D_{\Vec{\phi_l}} \}_l $ gates, which is sufficient to implement any single-qudit gate both for finite-field and integer-ring qudits \cite{Clark_2006, quditMBQC, quditMBQCstabilizerstates}. However, finite-field qudits allow for a more efficient decomposition of single-qudit gates into measurement patterns \cite{quditMBQCstabilizerstates}.

Furthermore, in Ref. \cite{quditMBQCstabilizerstates}, we have introduced resource states beyond graph states, where the qudits initialized in $\ket{+}$ are entangled with a more general block-diagonal two-qudit Clifford gate $G_E$. Then, an $X$ measurement on a two-qudit resource state results in the intrinsic single-qudit Clifford gate $G_I$ being applied. In odd prime-power dimensions, the overhead to decompose arbitrary single-qudit gates can then be lower than for the respective qudit cluster state resource with the intrinsic gate $H$.

For instance, qutrit resources, characterized by the ionic light-shift gate \cite{LightShiftGate}, allow for a decomposition of single-qutrit unitaries into measurement patterns with at most nine measurements on one-dimensional resource state chains, while on a cluster state chain, up to twelve measurements may be required when not modifying the intrinsic gate via adjusting measurement bases \cite{quditMBQCstabilizerstates}.

For universal quantum computing, the implementation of an entangling gate is necessary and sufficient \cite{QuditComputing, MathematicsQC, CriteriaQuditUniversality}. This is achieved using the existing vertical edges of the cluster state since a transport through an edge is equivalent to the entangling gate $CZ$ being applied. Control over where $CZ$ gates are applied is obtained by using $Z$ measurements to delete qudit vertices together with all attached edges and creating edges via $Y$ measurements. This is shown in Fig. \ref{fig:mbqc-basics} $(b)$.

The randomness of the single-qudit measurement outcomes appears as a Pauli by-product, which can be propagated to the end of the computation and accounted for in post-processing when the output is measured in the computational basis, since $Z(z)$ has no effect, and any $X(x)$ leads to re-interpreting the outcome via reversing the shift. The propagation of accumulated Pauli by-products on each logical qudit works because $H$ and $CZ$ are Clifford gates, diagonal gates commute with any $Z(z)$, and $X(k) D_{\Vec{\phi}} X(-k)$ remains diagonal. However, the latter means that, depending on previous measurement outcomes, diagonal gates have to be adjusted to the conjugated version (except for if the diagonal gate is itself Clifford).

\subsection{Blind quantum computing with qubits}
\label{subsec:blind-computing}

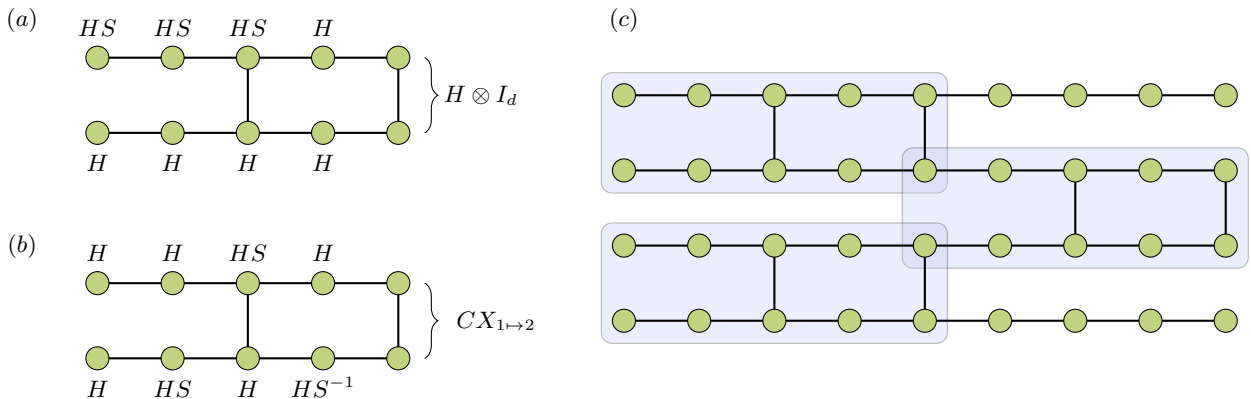
\begin{figure*}[t!]
    \centering
    \begin{tikzpicture}

    \node at (-3,0.5) {$(a)$};
    
    \draw[thick] (-2,0) -- (2,0);
    \draw[thick] (-2,-1) -- (2,-1);
    \draw[thick] (0,0) -- (0,-1);
    \draw[thick] (2,0) -- (2,-1);

    \foreach \x in {-2,...,2} {
        \foreach \y in {0,-1} {
            \node[circle, fill=pistacchio, draw=black] at (\x,\y) {}; 
        }
    }

    \draw [decorate,decoration={brace,amplitude=5pt,raise=10pt}] (2,0) -- (2,-1) node [black,midway,xshift=1.1cm] {$ H \otimes I_d $};

    \node at (-2,0.4) {$H S$};
    \node at (-2,-1.4) {$H$};
    
    \node at (-1,0.4) {$H S $};
    \node at (-1,-1.4) {$H$};

    \node at (0,0.4) {$H S$};
    \node at (0,-1.4) {$H$};
    
    \node at (1,0.4) {$H$};
    \node at (1,-1.4) {$H$};

    \node at (-3,-2.5) {$(b)$};

    \draw[thick] (-2,-3) -- (2,-3);
    \draw[thick] (-2,-4) -- (2,-4);
    \draw[thick] (0,-3) -- (0,-4);
    \draw[thick] (2,-3) -- (2,-4);

    \foreach \x in {-2,...,2} {
        \foreach \y in {-3,-4} {
            \node[circle, fill=pistacchio, draw=black] at (\x,\y) {}; 
        }
    }

    \draw [decorate,decoration={brace,amplitude=5pt,raise=10pt}] (2,-3) -- (2,-4) node [black,midway,xshift=1.3cm] {$ CX_{1 \mapsto 2}$};

    \node at (-2,0.4-3) {$H$};
    \node at (-2,-1.4-3) {$H$};
    
    \node at (-1,0.4-3) {$H$};
    \node at (-1,-1.4-3) {$H S$};

    \node at (0,0.4-3) {$H S$};
    \node at (0,-1.4-3) {$H$};
    
    \node at (1,0.4-3) {$H$};
    \node at (1,-1.38-3) {$H S^{-1}$};

    \node at (5,0.5) {$(c)$};

    \fill[periwinkle!80, opacity=0.3, draw=black, rounded corners=4pt] (4.7,0.3-0.5) rectangle (9.3,-1.3-0.5);
    \fill[periwinkle!80, opacity=0.3, draw=black, rounded corners=4pt] (8.7,-0.7-0.5) rectangle (13.3,-2.3-0.5);
    \fill[periwinkle!80, opacity=0.3, draw=black, rounded corners=4pt] (4.7,-1.7-0.5) rectangle (9.3,-3.3-0.5);

    \foreach \y in {-0.5,...,-3.5} {
            \draw [thick](5 , \y) -- (13, \y);
    }

    \draw[thick] (7,0-0.5) -- (7,-1-0.5);
    \draw[thick] (9,0-0.5) -- (9,-1-0.5);

    \draw[thick] (7,-2-0.5) -- (7,-3-0.5);
    \draw[thick] (9,-2-0.5) -- (9,-3-0.5);

    \draw[thick] (11,-1-0.5) -- (11,-2-0.5);
    \draw[thick] (13,-1-0.5) -- (13,-2-0.5);

    \foreach \x in {5,...,13} {
        \foreach \y in {-0.5,...,-3.5} {
            \node[circle, fill=pistacchio, draw=black] at (\x,\y) {}; 
        }
    }
\end{tikzpicture}
    \caption{Blind quantum computing with the qubit brickwork state \cite{Broadbent_2009}. On the elementary brickwork state unit, two logical qubits are processed, one along each horizontal resource state chain.
    The identity and any of the gates within $\{ H, S, T, CX \}$ can be realized via local measurements in the $X$ basis, rotated by angles in the finite-sized set $\mathcal{A}$ of Eq. $\eqref{eq:random-angles-qubits}$.
    $(a)$ Measurement pattern to implement the Hadamard gate. The single-qubit gates indicate the chosen measurement bases, where every Hadamard gate corresponds to an $X$ measurement while $HS$ corresponds to a measurement in the $X$ basis, rotated by $S^\dagger$.
    $(b)$ Implementation of the controlled-$X$ gate on an elementary qubit brickwork state unit.
    $(c)$ Arrangement of elementary units (highlighted) into the qubit brickwork state. Whenever logical qubits are not part of any elementary unit, they experience the identity, so transport without processing, if the respective physical qubits are measured in $X$ due to $H^2 = I_2$.
    }
    \label{fig:brickwork-patterns}
\end{figure*}

Blind quantum computing with qubits \cite{Broadbent_2009, introduction-blind-QC} relies on the measurement-based quantum computing framework. Here, an untrusted server prepares the resource state from qubits sent by the client and carries out the computation via single-qubit measurements without being able to retrieve any knowledge about the computation it performs, the input, or the output.

\subsubsection{Hiding single-qubit gates}

The first step to ensure server-blindness to the single-qubit gates being performed is for the client to rotate each qubit in $\ket{+}$ that composes the resource state via diagonal matrices $D_{\phi} = \textnormal{diag}(1,e^{i \phi})$ with random angles
\begin{equation}
    \phi \in \biggl\{ \frac{k \pi}{4} \bigm| k \in \{0,\hdots, 7 \} \biggl\} \eqcolon \mathcal{A},
    \label{eq:random-angles-qubits}
\end{equation}
which are kept secret \cite{Broadbent_2009, introduction-blind-QC}. Since $CZ$ commutes with diagonal gates, the server preparing the resource state from the rotated qubits is equivalent to first preparing the resource state and afterwards introducing random rotations.

The client and server communicate classically throughout, with the client instructing the server which single-qubit measurements to perform, taking into account the random rotation angles and Pauli by-product propagation.
The only single-qubit gates being performed measurement-based are from the set $\{ H, S, T \}$, where $S = D_{\frac{\pi}{2}}$ and $T = D_{\frac{\pi}{4}}$, which is sufficient to realize any single-qubit unitary with arbitrary precision since $H$ and $S$ generate the single-qubit Clifford group and $T$ is non-Clifford.

To implement the gate set $\{ H, S, T \}$, one either wants to measure in the $X$ basis without any rotation or with an $-\frac{\pi}{2}$ or $-\frac{\pi}{4}$ rotation. Which of these three options the client chooses is hidden to the server due to angles being randomized in multiples of $\frac{\pi}{4}$, Eq. \eqref{eq:random-angles-qubits}, so that irrespective of the client's measurement angle choice, the measurement angle distribution looks uniform to the server. The privacy of the single-qubit gates then essentially relies on employing a one-time pad. 
Usually, the client additionally randomly adds bit flips $D_{\pi}$, so that only the client can interpret the measurement outcome sequence. Hence, if the client wants to delegate a single-qubit measurement on the resource qubit $j$ in the $X$ basis, rotated by $D_{\phi}$, to the server, the client instructs to measure in the $X$ basis, rotated by $\phi + \phi_j + r_j \pi + \delta$, where $\phi_j \in \mathcal{A}$, $r_j \in \{0,1\}$ and $\delta$ being a potential Pauli by-product adjustment due to previous measurement outcomes.

\subsubsection{Hiding entangling gates}

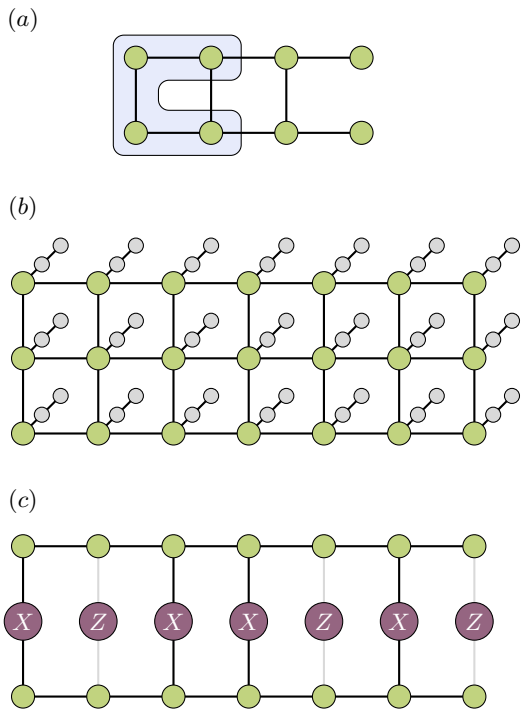
\begin{figure}[t!]
    \centering
    \begin{tikzpicture}

    \node at (6,0.5) {$(a)$};

    \draw[fill=periwinkle!30, draw=black, rounded corners=4pt] (7.5-0.3,0.3) -- (7.5+1.4,0.3) -- (7.5+1.4,-0.3) -- (7.5+0.3,-0.3) -- (7.5+0.3,-0.7) -- (7.5+1.4,-0.7) -- (7.5+1.4,-1.3) -- (7.5-0.3,-1.3) -- cycle;

    \draw[thick] (7.5,0) -- (10.5,0);
    \draw[thick] (7.5,-1) -- (10.5,-1);

    \draw[thick] (7.5,0) -- (7.5,-1);
    \draw[thick] (8.5,0) -- (8.5,-1);
    \draw[thick] (9.5,0) -- (9.5,-1);

    \foreach \x in {7.5,...,10.5} {
        \foreach \y in {0,-1} {
        \node[circle, fill=pistacchio, draw=black] at (\x,\y) {};
        }
    }

    \node at (6,-2) {$(b)$};

    \foreach \x in {6,...,12} {
        \draw[thick] (\x,-3) -- (\x,-5);
    }
    
    \foreach \y in {-3,...,-5} {
        \draw[thick] (6,\y) -- (12,\y);
    }

    \foreach \x in {6,...,12} {
        \foreach \y in {-3,...,-5} {
        \draw[thick] (\x,\y) -- (\x+0.5,\y+0.5);
        \node[circle, fill=black!15, draw=black, minimum size=2mm, inner sep=0pt] at (\x + 0.25, \y +0.25) {};
        \node[circle, fill=black!15, draw=black, minimum size=2mm, inner sep=0pt] at (\x + 0.5, \y +0.5) {};
        \node[circle, fill=pistacchio, draw=black] at (\x,\y) {};
        }
    }

    \node at (6,-5.9) {$(c)$};
      
    \draw[thick] (6,-6.5) -- (12,-6.5);
    \draw[thick] (6,-8.5) -- (12,-8.5);

    \foreach \x in {6,8,9,11} {
        \draw[thick] (\x, -6.5) -- (\x, -8.5) ;
        \node[circle, fill=mutedplum!80, draw=black, text=white, minimum size=5mm, inner sep=0pt] at (\x,-7.5) {$X$}; 
    }

    \foreach \x in {7,10,12} {
        \draw[thick, opacity=0.15] (\x, -6.5) -- (\x, -8.5) ;
        \node[circle, fill=mutedplum!80, draw=black, text=white, minimum size=5mm, inner sep=0pt] at (\x,-7.5) {$Z$}; 
    }

    \foreach \x in {6,...,12} {
        \foreach \y in {-6.5,-8.5} {
            \node[circle, fill=pistacchio, draw=black] at (\x,\y) {}; 
        }
    }
\end{tikzpicture}
    \caption{To hide the positions of entangling gates from the server, different strategies, corresponding to different resource state architectures, can be chosen beyond the qubit brickwork state in Fig. \ref{fig:brickwork-patterns}.
    $(a)$ The open-ended cluster state allows for universal quantum computing without unhidden and, thus, structure-revealing $Z$ deletion measurements \cite{clusteruniversality-XYmeasurements}. Here, the highlighted unit corresponds to $C_2$ in Eq. \eqref{eq:layer-operator} if the qubits in the first column are measured in $X$.
    $(b)$ Hair implantation technique, where each cluster state qubit is decorated with a two-qubit chain (light grey) that allows for simulating both rotated $X$ and $Z$ basis measurements without performing $Z$ measurements. This naturally generalizes to graph states, including cluster states in higher dimensions. $(c)$ Graph hiding, where ancillary qubits (dark violet) are initialized in the $X$ or $Z$ basis (white text), which cannot be distinguished by the server, depending on where entangling gates should be placed. The controlled-phase $CZ$ gate then does not have any effect on the $Z$ basis qubits (greyed out edges) while $X$ basis qubits are entangled with the resource. Similar structures are referred to as the square brickwork state in Ref. \cite{ImprovedBrickworkState}.
    }
    \label{fig:hiding-entangling-gate}
\end{figure}

The application of entangling gates should also remain private, which is not the case if the client instructs the server for $Z$ deletion measurements, such as in Fig. \ref{fig:mbqc-basics} $(b)$, since these measurements can not be hidden with the previously introduced technique of randomly applying single-qubit rotations.

Therefore, in the original proposal for blind quantum computing with qubits \cite{Broadbent_2009}, the brickwork state was introduced as a resource.
The elementary unit of a brickwork state supports the measurement-based implementation of all diagonal gates, the Hadamard gate, and the controlled-$X$ gate $CX$ with only rotated $X$ basis measurements, as displayed in Figs. \ref{fig:brickwork-patterns} $(a)$ and $(b)$, respectively.
Diagonal gates $D_{\phi}$ can be implemented by rotating the $X$ basis by $D_{\phi}^\dagger$ during the first measurement on either of the two logical qubits.
The arrangement of elementary units into the brickwork state resource, allowing for $CX$ gates between arbitrary neighboring logical qubits, is shown in Fig. \ref{fig:brickwork-patterns} $(c)$.

As discussed in the following, other hiding strategies have been proposed associated with different resource state architectures, which are summarized in Fig. \ref{fig:hiding-entangling-gate}.

\paragraph{The open-ended cluster state}

It turns out that even when prohibiting unhidden structure-revealing $Z$ deletion measurements, the qubit cluster state is universal \cite{clusteruniversality-XYmeasurements}.
To understand cluster state quantum computing without deletion measurements, we first study the effect of performing $X$ measurements on all qubits except the last output column. Herein, we consider an open-ended cluster state \cite{clusteruniversality-XYmeasurements}, where the output qubits are not connected via vertical entangling gates.

Since an $X$ measurement on an isolated horizontal chain implements a Hadamard gate and each vertical edge a $CZ$ gate, performing $X$ measurements on all qubits of one column implements the operator
\begin{equation}
    C_n \coloneq \prod_{j=1}^n H_j \prod_{j=1}^{n-1} CZ_{j,j+1}.
    \label{eq:layer-operator}
\end{equation}

Considering a qubit open-ended cluster state of lattice size $n \cross (n+2)$, it was shown that $X$ measurements on all qubits except the output, implementing $n+1$ times the operator $C_n$, act as a global mirror, reflecting each qubit state along the intermediate horizontal axis \cite{clusteruniversality-XYmeasurements, MirrorOperator1, MirrorOperator2}. In Fig. \ref{fig:hiding-entangling-gate} $(a)$, an example for $n=2$ logical qubits is shown, where the mirror corresponds to a swap gate being performed.

By introducing rotated $X$ basis measurements at different positions of the open-ended cluster state, it was subsequently shown how to implement single-qubit gates and an entangling gate between neighboring pairs of logical qubits \cite{clusteruniversality-XYmeasurements}.

In particular, a rotation of the $X$ basis in the first column of the open-ended cluster state realizes a diagonal gate $D_{\phi}$, so a $Z$ rotation $e^{-i \frac{\phi}{2} Z}$, on either of the logical qubits while a rotation in the $n+1$-th column implements $H D_{\phi} H^\dagger $, an $X$ rotation $e^{-i \frac{\phi}{2} X}$. Single-qubit rotations around these two axes suffice to decompose any single-qubit unitary \cite{NielsenChuang}.

A two-qubit entangling gate between arbitrary neighboring logical qubit pairs on the open-ended cluster state is realized in Ref. \cite{clusteruniversality-XYmeasurements} by rotating the $X$ measurement by $\phi$ in the first (or last) row and column $m$ with $1 < m < n+1$, observing that repeated conjugation via $C_n$ yields
\begin{equation*}
    D_{\phi} \propto e^{- i \frac{\phi}{2} Z_1} \xmapsto{C_n} e^{- i \frac{\phi}{2} X_1} \xmapsto{C_n} e^{- i \frac{\alpha}{2} X_1 Z_2} \xmapsto{C_n} e^{- i \frac{\phi}{2} X_2 Z_3}.
\end{equation*}
Each application of $C_n$ shifts the Pauli string $X_1 Z_2$ to the next pair of qubits. Hence, the qubit entangling gate $e^{-i \frac{\phi}{2} X_k \otimes Z_{k+1}}$ can be steered to act at an arbitrary position $k \in \{1,\hdots, n-1 \}$.

\paragraph{The decorated cluster state}

Alternatively, the hair implantation technique \cite{BlindTopologicalMBQC} has been proposed to hide deletion measurements.

Here, each cluster state qubit is decorated with a two-qubit chain, as shown in Fig. \ref{fig:hiding-entangling-gate} $(b)$, allowing us to simulate the effect of a $Z$ deletion measurement with only rotated $X$ measurements as well as the effect of a rotated $X$ measurement. Then, one can carve out the desired cluster state structure similarly to Fig. \ref{fig:mbqc-basics} $(b)$ without using unhidden $Z$ measurements.

In particular, the effect of a $Z$ measurement on any cluster state qubit is simulated by measuring the cluster state qubit and both hair qubits in the $X$ basis, rotated by $S$, starting with the cluster state qubit and moving from there along the hair \cite{BlindTopologicalMBQC}.

Instead, the effect of an $X$ basis measurement, rotated by $D_{\phi}$, on the cluster state qubit can be simulated by measuring the cluster state qubit and first hair qubit in $X$ while the second hair qubit is measured in the $X$ basis, rotated by $D_{\phi}$ \cite{BlindTopologicalMBQC}.

\paragraph{Graph hiding}

Another ancillary-assisted way for graph hiding \cite{GraphHiding, ImprovedBrickworkState} is to use qubits, which either bridge, so entangle, two horizontal chains processing a logical qubit each, or break, so disentangle, them. An example of such graph hiding is shown in Fig. \ref{fig:hiding-entangling-gate} $(c)$.

Depending on whether one wants to bridge or break, so whether the entangling gate $CZ$ is supposed to have an effect, the ancillary qubits are either initialized randomly in the $X$ basis, within the set
\begin{equation*}
    \{ \ket{+}, \ket{-} \} = \{ \ket{0_X}, \ket{1_X} \} = \{ H \ket{0}, H \ket{1} \}, 
\end{equation*}
or in the $Z$ basis $\{ \ket{0}, \ket{1} \}$.
The privacy of the graph state structure then relies on the measurement bases $\{ \ket{+}, \ket{-} \}$ and $\{ \ket{0}, \ket{1} \}$ being indistinguishable for the server.

In Ref. \cite{ImprovedBrickworkState}, this graph hiding strategy was used to define variants of the qubit brickwork state. For the square brickwork state \cite{ImprovedBrickworkState}, connectivity between all neighboring logical qubits via vertical edges is given, similar to Fig. \ref{fig:hiding-entangling-gate} $(c)$.
In the hyper-brickwork state \cite{ImprovedBrickworkState}, all logical qubits are connected with each other via ancillary qubits, initialized in the basis that determines whether a bridge or a break happens. This increases the connectivity since entangling gates can then also be performed between non-neighboring logical qubits at the cost of increasing the number of ancillae. For states that only permit nearest-neighbor interactions, such a structure is not possible, which is why the circular brickwork state has been introduced as a further alternative \cite{ImprovedBrickworkState}.

\subsubsection{Identifying dishonest server behavior}

To identify potentially dishonest behavior from the server, multiple verification techniques have been proposed.
For instance, the client may designate certain logical qubits as traps, whose expected outcomes are computed in advance. These traps allow the client to detect any deviations by the server from the prescribed protocol \cite{introduction-blind-QC}. 
This works for all the introduced resource state structures.

For the decorated cluster state, instead of logical qubits, individual physical qubits can be chosen as traps, disentangling them from qubits on neighboring sites of the resource via simulated $Z$ deletion measurements.
Since the client knows in advance what measurement outcomes these traps should produce, any discrepancy reveals a dishonest run of the intended protocol \cite{introduction-blind-QC}.

Moreover, in the measurement-only client setting \cite{BQCmeasurementonlyclient-first}, where the client receives the resource qubits (either directly or teleported through a successfully distributed Bell pair) to perform the measurement-based computation themself, the client may introduce resource graph state verification \cite{VerificationMeasurementOnlyBQC, BQCmeasurementonlyclient, BQCmeasurementonly-inputverification}.

\section{Qudit blind quantum computing}
\label{sec:qudit-blindQC}

There are two steps to generalizing blind quantum computing to qudits. First, one needs to find a universal single-qudit gate set and show that this set can be implemented blindly, ensuring privacy, on the resource state. This is discussed in Sec. \ref{subsec:prime-power-universal}. Second, one needs to demonstrate hiding strategies for a two-qudit entangling gate, Sec. \ref{subsec:hiding-entangling-gates}. For each resource state architecture, we discuss associated overheads and strategies for identifying dishonest server behaviour. Finally, we outline advantages of utilizing qudits instead of qubits in Sec. \ref{subsec:qudit-advantages}.

\subsection{Hiding single-qudit gates}
\label{subsec:prime-power-universal}

As for qubits, the qudit entangling gate $CZ$ commutes with diagonal single-qudit gates $D_{\Vec{\phi}}$. Hence, the client can randomly apply diagonal unitaries to resource qudits, keeping the angles $\{\phi_k\}_k$ of each rotation private before the qudits are sent to the server and entangled to the corresponding resource state. Accounting for the angles when instructing the server for measurements in rotated $X$ bases, the implemented single-qudit gate remains private due to the client imitating a uniform angle distribution to the server.

In the original approach with qubits \cite{Broadbent_2009}, the angles are randomly chosen from the set $\mathcal{A}$, Eq. \eqref{eq:random-angles-qubits}, allowing to implement any operation of the approximately universal single-qubit gate set $\{H, S, T \}$ blindly. Since for finite-field qudits the Clifford group becomes approximately universal when supplementing any non-Clifford single-qudit gate \cite{heinrich2021stabiliser} (note that this is not established for integer-ring qudits of arbitrary dimension), we can follow the same approach for prime-power-dimensional qudits and need to find one non-Clifford single-qudit diagonal gate.

Generalizing the approach of approximately universal gate sets to finite-field qudits, we search for an analogue of the qubit $T$ gate, which is motivated by evidence that $T$ gates in prime dimensions are maximally robust to depolarizing and phase-damping noise in analogy with the qubit case \cite{generalizedTgate}.
For prime dimensions, the $T$ gate is generalized by observing that the qubit $T$ gate conjugates $X$ to $XS$ up to a phase. In dimension $d=3$, it is then given by \cite{generalizedTgate}
\begin{equation}
    T_3 = \left( \begin{array}{ccc}
    1 & 0 & 0 \\
    0& e^{2\pi i/9} & 0 \\
    0 & 0 & e^{-2\pi i/9}
    \end{array} \right).
\label{eq:qutrit-Tgate}
\end{equation}
For $d> 3$, one can pick \cite{generalizedTgate, generatorsTgate2, quditcolorcode-Tgate, BenchmarkingUniversalQuditGateset}
\begin{equation*}
    T_{d} = \sum_{k} \omega_{d}^{k^3 6^{-1}} \ket{k} \bra{k},
\end{equation*}
where $\omega_d = e^{\frac{2 \pi i}{d}}$. The reason that $T_d$ is qualitatively different from both $T$ and $T_3$ can be traced back to the integer six not being invertible in $\mathbb{Z}_2$ and $\mathbb{Z}_3$ \cite{generatorsTgate2}.

For finite-field qudits of prime-power dimension $p^m$ with $p \notin \{2,3 \}$, we can then pick 
\begin{equation}
    T_d^F = \sum_k \chi(6^{-1} k^3) \ket{k} \bra{k},
    \label{eq:T-other-prime-powers}
\end{equation}
as discussed in Appendix \ref{app:nonclifford-diagonal}.
However, for $p \in \{2,3 \}$, this gate is clearly Clifford since then $k^3 = k$.
To lift the qubit $T$ gate to even prime-power dimensions with $p=2$, we, therefore, instead define
\begin{equation}
    T_{2^m}^F \coloneq \sum_k \chi_8(k^4) \ket{k} \bra{k},
    \label{eq:even-power-T}
\end{equation}
where $\chi_8(t) = \omega_8^{\tr_8(t)}$ and the multiplication map trace $\tr_8(t)$ is computed in the Galois ring $\mathbb{GR}_{8^m}$ and takes values in $\mathbb{Z}_8$.

For prime-power dimensions with $p=3$, we take
\begin{equation}
    T_{3^m}^F \coloneq \sum_k \chi_9(k^3) \ket{k} \bra{k},
    \label{eq:three-power-T}
\end{equation}
where $\chi_9(t) = \omega_9^{\tr_9(t)}$ and $\tr_9(t)$ is evaluated in the Galois ring $\mathbb{GR}_{9^m}$, taking values in $\mathbb{Z}_9$. For $m=1$, this definition coincides with the $T_3$ gate.

In Appendix \ref{app:nonclifford-diagonal}, we show by conjugation of $X(x)$ to non-Pauli Clifford gates that these generalized $T$ gates are indeed diagonal non-Clifford operators.

Hence, for even prime-power dimensions, $p=2$, we can keep the angle set $\mathcal{A}$ of Eq. \eqref{eq:random-angles-qubits} to implement approximately universal gate sets server-blindly. For $p=3$, we replace it via
\begin{equation*}
    \biggl\{ \frac{2 \pi k}{9} \bigm| k \in \{0,\hdots, 8 \} \biggl\},
\end{equation*}
whereas for the remaining prime-power dimensions with $p \notin \{2,3\}$, we take
\begin{equation*}
    \biggl\{ \frac{2 \pi k}{p} \bigm| k \in \{0,\hdots, p-1 \} \biggl\}.
\end{equation*}
Introducing random rotations on the resource qudits sent to the server with angles from these respective sets and instructing for the desired measurement bases that account for these angles, the client then maintains privacy for all executed single-qudit gates.

Furthermore, the Solovay-Kitaev theorem \cite{solovay-kitaev} then states that there is a constant $c$ such that any special unitary can be approximated with accuracy $\epsilon$ via a sequence from an approximately universal gate set of length $O(\log^c(1/\epsilon))$. The Solovay-Kitaev algorithm \cite{solovay-kitaev} yields the approximation sequence for qudits with $c \approx 4$ independent of the dimension. For prime-power-dimensional finite-field qudits and exact universality relying on the availability of arbitrary diagonal unitaries, we discuss in Ref. \cite{quditMBQCstabilizerstates} that the number of measurements to implement any single-qudit gate along a one-dimensional cluster state chain scales linearly with the dimension of the qudit system.

If we allow uniform sampling and communicating values from the continuous interval $[0,2 \pi]$, we have an exactly universal gate set in arbitrary dimension, both for integer-ring and finite-field qudits at our disposal \cite{quditMBQC,quditMBQCstabilizerstates}. Thus, we could do blind quantum computing in any finite dimension.
However, the ability of exact decomposition comes at the cost of increased difficulty in sampling and classically communicating quasi-continuous numbers to the server.

Moreover, this hiding technique of single-qudit gates also works for qudit resource states, characterized by diagonal Clifford entangling gates $G_E$ other than $CZ$ \cite{quditMBQCstabilizerstates} since diagonal entangling gates commute with all other diagonal gates.
As for the cluster or brickwork state resources, the client then randomly applies diagonal single-qudit gates to the resource qudits, initially in $\ket{0_X}$, before they are entangled via $G_E$ by the server. This mimics a uniform angle distribution to the server, irrespective of which rotated $X$ measurement basis the client chooses.

\subsection{Hiding entangling gates}
\label{subsec:hiding-entangling-gates}

After discussing the privacy of single-qudit gates in the previous section, we now investigate how entangling gates can be hidden, covering the brickwork state, the open-ended and decorated cluster states, and, finally, graph states and qudit resource state variants beyond.

Since we have seen that only measurements in rotated $X$ bases can be hidden from the server, only these single-qudit measurement bases are allowed in each resource state architecture, associated with a different entangling gate hiding strategy. In particular, $Z$ deletion measurements, which cannot be kept private and would reveal information about the computation, are prohibited.

\subsubsection{The qudit brickwork state}
\label{subsubsec:qudit-brickwork}

\begin{figure}[t!]
    \centering
    \begin{tikzpicture}
    \draw[thick] (-2,0) -- (2,0);
    \draw[thick] (-2,-1) -- (2,-1);
    \draw[thick] (0,0) -- (0,-1);
    \draw[thick, dashed] (2,0) -- (2,-1);

    \foreach \x in {-2,...,2} {
        \foreach \y in {0,-1} {
            \node[circle, fill=pistacchio, draw=black] at (\x,\y) {}; 
        }
    }
\end{tikzpicture}
    \caption{Qudit brickwork state elementary unit, where the dashed edge corresponds to $CZ^{-1}$. The elementary units can be arranged into the qudit brickwork state in the same manner as for qubits, Fig. \ref{fig:brickwork-patterns} $(c)$. Each elementary unit supports the realization of any diagonal gate, the Hadamard gate, or the $CX$ entangling gate on the two logical qudits.
    }
    \label{fig:brickwork-resource}
\end{figure}
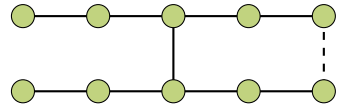

The elementary unit of the qubit brickwork state \cite{Broadbent_2009} is displayed in Figs. \ref{fig:brickwork-patterns} $(a)$ and $(b)$. Compared to the open-ended cluster state resources, this unit has a fixed size of ten qubits and, in addition, requires fewer entangling gates, leading to higher fidelities of measurement-induced gates.

To generalize the brickwork state to qudits, we replace the last vertical edge in its elementary unit with $CZ^{-1}$, drawn as a dotted edge in Fig. \ref{fig:brickwork-resource}. The arrangement of the elementary units into the qudit brickwork state is then analogous to the qubit case, see Fig. \ref{fig:brickwork-patterns} $(c)$. As for qubits, the two logical qudits, not participating in any elementary unit experience the identity if they are measured in $X$ due to $H^4 = I_d$ in all dimensions $d$, both for finite-field and integer-ring qudits.

If we measure all qudits of the modified elementary brickwork state unit, Fig. \ref{fig:brickwork-resource}, except the output in the $X$ basis, we realize the identity gate due to the effectively executed quantum circuit being (ignoring a potential Pauli by-product)
\begin{equation*}
\begin{aligned}
    & CZ^{-1} (H^2 \otimes H^2) CZ (H^2 \otimes H^2) 
    \\ & = CZ^{-1} (M(-1) \otimes M(-1)) CZ (M(-1) \otimes M(-1)) 
     \\ & = CZ^{-1} CZ  = I_d.
\end{aligned}
\end{equation*}

If we rotate the $X$ basis by $D_{\Vec{\phi}}^\dagger$ on either of the two logical qudits during the first measurement, we apply $D_{\Vec{\phi}}$ measurement-based.

Furthermore, we can implement the Hadamard gate on either of the two logical qudits by using the qudit analogue of the measurement pattern in Ref. \cite{Broadbent_2009}, displayed in Fig. \ref{fig:brickwork-patterns} $(a)$. Here, the first three $X$ basis measurements are rotated by $S(1)^\dagger$, so $S^\dagger$ for even prime-power dimensions. We show that these measurements realize a Hadamard gate in Appendix \ref{app:brickwork-Hadamard}. For universal quantum computing, the realization of an entangling gate is missing. With the measurement pattern in Ref. \cite{Broadbent_2009}, Fig. \ref{fig:brickwork-patterns} $(b)$, generalized to qudits, we realize a $CX$ gate, as demonstrated in Appendix \ref{app:brickwork-CX}.

If we do not have access to the entangling gate $CZ^{-1}$ or want to keep all entangling gates equal to $CZ$ in the brickwork state resource, we can use that $CZ^{p-1} = CZ^{-1}$ (for integer-ring qudits, Appendix \ref{app:integer-ring-qudits}, it is $CZ^{d-1} = CZ^{-1}$). The server can either apply $CZ$ multiple times to obtain $CZ^{-1}$ or the elementary unit could be modified, replacing the dashed edge in Fig. \ref{fig:brickwork-resource} with $CZ$ and adding $p-2$ further ten-qudit blocks that each implement $CZ$ (for instance, the analogue of the qubit elementary unit without the first or last vertical edge).
However, this has the disadvantage that the size of each elementary brickwork state unit then scales with $p$, so the dimension of the system $d=p^m$.

As for qubits, dishonest behaviour of the server can be identified by using some of the logical qudits as traps. Here, the client pre-computes the measurement outcome on each trap qudit, which would be obtained in an honest run of the protocol, so that deviations from the server can be detected.

\subsubsection{The open-ended cluster state}
\label{subsubsec:qudit-clusterstate-resource}

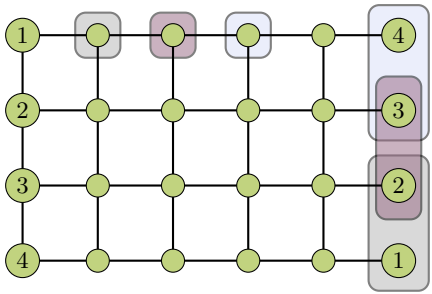
\begin{figure}[t!]
    \centering
    \begin{tikzpicture}

    \fill[black!30, opacity=0.5, draw=black, thick, rounded corners=4pt] (6.7,-3.3) rectangle (7.3,-2.7);
    \fill[black!30, opacity=0.5, draw=black, thick, rounded corners=4pt] (10.6,-6.4) rectangle (11.4,-4.6);

    \fill[periwinkle!50, opacity=0.5, draw=black, thick, rounded corners=4pt] (8.7,-3.3) rectangle (9.3,-2.7);
    \fill[periwinkle!50, opacity=0.5, draw=black, thick, rounded corners=4pt] (10.6,-4.4) rectangle (11.4,-2.6);

    \fill[mutedplum!80, opacity=0.5, draw=black, thick, rounded corners=4pt] (7.7,-3.3) rectangle (8.3,-2.7);
    \fill[mutedplum!80, opacity=0.5, draw=black, thick, rounded corners=4pt] (10.7,-5.45) rectangle (11.3,-3.55);

    \foreach \x in {6,...,10} {
        \draw[thick] (\x,-3) -- (\x,-6);
    }
    
    \foreach \y in {-3,...,-6} {
        \draw[thick] (6,\y) -- (11,\y);
    }

    \foreach \x in {7,...,10} {
        \foreach \y in {-3,...,-6} {
        \node[circle, fill=pistacchio, draw=black] at (\x,\y) {};
        }
    }

    \foreach \y/\label in {-3/1, -4/2, -5/3, -6/4} {
    \node[circle, fill=pistacchio, minimum size=4.5mm, inner sep=0pt, draw=black]
       at (6,\y) {\label};
    }
    
    \foreach \y/\label in {-3/4, -4/3, -5/2, -6/1} {
    \node[circle, fill=pistacchio, minimum size=4.5mm, inner sep=0pt, draw=black]
       at (11,\y) {\label};
    }
    
\end{tikzpicture}
    \caption{Elementary unit of lattice size $4 \cross 6$ for the open-ended qudit cluster state and $n=4$ logical qudits.
    The numbers indicate the flow of information, the logical qudits, from the input on the left towards the output on the right if $X$ measurements are performed on all qudits except the output due to the qudit mirror being performed.
    Measurements in the first column serve to implement diagonal gates on the input. Instead, measurements in the fifth column allow the measurement-based implementation of diagonal gates, conjugated by the Hadamard gate, on the output qudits. Rotated $X$ bases measurements in the first row and a column between the first and fifth allow us to realize entangling gates between different pairs of output qudits, chosen by the column, as indicated in the figure.
    }
    \label{fig:openended-clusterstate}
\end{figure}

To understand cluster state quantum computing without deletion measurements, we first study the effect of performing $X$ measurements on all qudits except the output, as in the qubit variant. Herein, we consider an open-ended cluster state \cite{clusteruniversality-XYmeasurements} unit of lattice size $n \cross (n+2)$ for $n$ logical qudits, as shown in Fig. \ref{fig:hiding-entangling-gate} $(a)$ for $n=2$, where the output qudits in the last column are not connected via vertical entangling gates.

Even though the size of an elementary unit is rather large since its depth increases linearly with the number of logical qudits, cluster states are highly symmetric and a natural choice for some experimental platforms with nearest-neighbor interactions.

In Appendix \ref{app:qudit-mirror}, we prove that $n+1$ applications of $C_n$, Eq. \eqref{eq:layer-operator} with the Hadamard and controlled-phase gates generalized to qudits, describes a global mirror, reflecting each logical qudit along the middle horizontal axis, if $n$ is even and a global mirror, supplemented by $M(-1) = H^2$, with $M(-1)$ from Eq. \eqref{eq:multiplication-gate}, on every qudit for $n$ odd. Note that in even prime-power dimensions, such as for qubits, $M(-1)$ is the identity.

If the delegated computation is supposed to happen on an odd number of logical qudits, one can always use excess qudits as further trap qudits for the server to keep $n$ even and have the complete analogue of the qubit mirror on qudits, avoiding the additional $M(-1)^{\otimes n}$ operation.

In analogy to Ref. \cite{clusteruniversality-XYmeasurements}, we now consider which gates are executed if one of the $X$ measurements in the first or last column of the open-ended cluster state is rotated.

For instance, measuring a qudit of the first column in the $X$ basis with a rotation $D_{\Vec{\phi}}^\dagger$ is equivalent to $D_{\Vec{\phi}}$ being performed before the global qudit mirror acts due to \begin{equation*}
    (H D_{\Vec{\phi}}  \otimes I) CZ = (H   \otimes I) CZ (D_{\Vec{\phi}} \otimes I_d).
\end{equation*}
When a qudit in the last non-output column is instead rotated by $D_{\Vec{\phi}}^\dagger$, it holds that
\begin{equation*}
    (H D_{\Vec{\phi}}  \otimes I) CZ = (H D_{\Vec{\phi}} H^\dagger  \otimes I) (H \otimes I) CZ,
\end{equation*}
so that effectively $H D_{\Vec{\phi}} H^\dagger$ is realized after the global mirror operator.

To implement the Hadamard gate via diagonal and conjugated diagonal unitaries, we use the multiplication gate decomposition of Eq. \eqref{eq:multiplication-gate-decomposition}, observing that
\begin{equation*}
    M(-1) = H^2 = H S(-1) H S(-1) H S(-1).
\end{equation*}
This can be rewritten via
\begin{equation*}
\begin{aligned}
    & H =  S(-1) H S(-1) H^\dagger M(-1) S(-1)
    \\ & =  S(-1) H S(-1) H^\dagger S(-1) M(-1),
\end{aligned}
\end{equation*}
implying that $H^{-1} =  S(-1) H S(-1) H^\dagger S(-1)$ and
\begin{equation*}
\begin{aligned}
    & H = S(1) H S(1) H^\dagger S(1).
\end{aligned}
\end{equation*}
Hence, in addition to arbitrary diagonal gates, we are also able to perform the Hadamard gate measurement-based.

To implement a qudit entangling gate, we now consider rotated measurements in columns between the first and second last (the last non-output column).
We observe in Appendix \ref{app:qudit-mirror} that the Pauli string $Z_1^{-1} X_2$ shifts with each application of $C_n$ as for qubits and, in addition, the powers of the Pauli string alternate.
However, since $Z$ is no longer Hermitian for qudits besides in even prime-power dimensions, an operator such as $e^{i \alpha Z}$ is not a qudit unitary in general.

Still, performing measurement-based the diagonal gate
\begin{equation*}
    D_{\Vec{\alpha}} = e^{i \sum_k \alpha_k \ket{k_Z} \bra{k_Z}}
\end{equation*}
in the first row and a column $m \in \{2,\hdots, n\}$, we demonstrate in Appendix \ref{app:entangling-gate-clusterstate} that repeated conjugation by $C_n$ (in total $n+2 - m$ times) applies the entangling gate
\begin{equation*}
    e^{i \sum_{k,j} \alpha_k \ket{-(k+j)_Z} \bra{(k+j)_Z} \otimes \ket{j_X} \bra{j_X}}
\end{equation*}
to the output qudits $n+1-m$ and $n+2 - m$ (which correspond to the input qudits $m$ and $m-1$) for $n+2-m$ even. For $n+2-m$ odd, instead the entangling gate
\begin{equation*}
\begin{aligned}
    & e^{i \sum_{k,j} \alpha_k \ket{(k-j)_Z} \bra{(k-j)_Z} \otimes \ket{j_X} \bra{j_X}}
\end{aligned}
\end{equation*}
is applied. Hence, if we want to apply an entangling gate to qudits $j$ and $j-1$ of the input, we can rotate the $X$ basis measurement on the qudit in the first row and column $m = j$.
We visualize the column choice in Fig. \ref{fig:openended-clusterstate} in the case of four logical qudits.

To be able to propagate Pauli by-products, we select the angles of $D_{\Vec{\alpha}}$ such that we obtain a Clifford entangling gate. For this, we can choose $D_{\Vec{\alpha}}$ as a local Clifford gate, such as one of the phase gates in $\{ S(\lambda) \}_{\lambda \in \mathbb{F}_d}$.
Since we then have conjugated a single-qudit Clifford gate by the Clifford operators $CZ$ and $H$ in $C_n$, the resulting entangling gate has to be a two-qudit Clifford gate.

\subsubsection{The decorated qudit cluster state}
\label{subsubsec:qudit-decorated-clusterstate}

Another idea to avoid unhidden $Z$ deletion measurements is to use a decorated cluster state \cite{BlindTopologicalMBQC}, where to each qudit a string of two ancillary qudits, a hair, is attached, Fig. \ref{fig:hiding-entangling-gate} $(b)$. In the qubit case, the ancillae can then be used to simulate both a $Z$ and a rotated $X$ basis measurement on the cluster state qubits without physically performing $Z$ measurements. Hence, the type of measurement can be hidden from the server. This strategy works for arbitrary graph states, so one could also use it for the three-dimensional cluster state \cite{BlindTopologicalMBQC}, for instance.

To understand the measurement simulation for qudits, we consider two ancillae, initially in the state $\ket{+}^{\otimes 2}$, attached to a qudit of the resource state in an arbitrary state $\ket{\psi}$.
After an $X$ measurement on the resource qudit with outcome $k_1$ and a subsequent $X$ measurement with outcome $k_2$ on the first hair qudit, the second hair qudit is in the state
\begin{equation*}
\begin{aligned}
    & H Z(-k_2) H Z(-k_1) \ket{\psi}
    = X(k_2) Z(k_1) H^2 \ket{\psi}.
\end{aligned}
\end{equation*}
The last $X$ measurement with outcome $k_3$ then yields
\begin{equation*}
\begin{aligned}
    & \bra{(k_3)_X} X(k_2) Z(k_1) H^2 \ket{\psi}
    \propto \bra{(k_3-k_1)_X} H^2 \ket{\psi}
    \\ & = \bra{(k_3-k_1)_X} M(-1) \ket{\psi} = \bra{(k_1-k_3)_X} \ket{\psi}.
\end{aligned}
\end{equation*}
This scenario is then equivalent to an $X$ measurement of the cluster state qudit with outcome $k_1 - k_3$ (the outcome $k_2$ of the second $X$ measurement just introduces an irrelevant global phase).
Hence, we can simulate the effect of an $X$ measurement on the resource qudit.

If we rotate the $X$ basis during the first measurement, we can simulate the effect of a diagonal gate, for instance, one of the phase gates in $\{ S(\lambda \})_{\lambda \in \mathbb{F}_d}$ or the qudit $T$ gate, followed by an $X$ measurement.

Rotating instead the $X$ basis by $S(1)^\dagger$ (for even prime-power dimensions $S^\dagger$) for all three measurements, we implement a $Z$ measurement since
\begin{equation*}
\begin{aligned}
    & \bra{(k_3)_X} S(1) H S(1) Z(- k_2) H S(1) Z(- k_1) \ket{\psi}
    \\ & = \bra{(-k_3)_Z} H S(1) X(k_2) H S(1) X(k_1) H S(1) \ket{\psi}
    \\ & \propto \bra{(-k_3)_Z} X(-k_2) Z(k_2 - k_1) H S(1) H S(1) H S(1) \ket{\psi}
    \\ & = \bra{(k_2 - k_3)_Z} \ket{\psi},
\end{aligned}
\end{equation*}
where we use $H^{-1} = H^3 = M(-1) H$ and the identity decomposition $I_d = M(1) = H S(1) H S(1) H S(1)$, following from Eq. \eqref{eq:multiplication-gate-decomposition}. In total, we simulate the effect of a $Z$ deletion measurement with outcome $\ket{(k_2 - k_3)_Z}$.

Thus, we can simulate all required measurements on the cluster state with only rotated $X$ basis measurements that can be kept private.

\subsubsection{Graph hiding}
\label{subsubsec:graph-hiding}

One way to hide the geometry of a resource graph state was introduced in the previous section for the decorated cluster state, Fig. \ref{fig:hiding-entangling-gate} $(b)$, where two ancillary qudits, a hair, were attached to each resource qudit to simulate all required measurements without revealing the measurement choice to the server.

The graph hiding technique in Fig. \ref{fig:hiding-entangling-gate} $(c)$ directly generalizes to qudit graph states. To keep the positioning of entangling gates private, ancillary qudits are initialized in the $X$ basis if the entangling gate $CZ$ is supposed to have an effect and in the $Z$ basis if not. Since the $X$ and $Z$ bases are indistinguishable, the server cannot infer any information about the graph state structure that it prepares.
Measurement-only clients may use resource graph state verification to identify malicious server behaviour, for which protocols have been generalized to qudit graph states of local prime dimension \cite{VerificationQuditGraphStatesPrimeDimension}.

Graph hiding also applies to qudit resources, characterized by diagonal Clifford entangling gates $G_E$ other than $CZ$ \cite{quditMBQCstabilizerstates}, since also $G_E$ creates no entanglement when applied to qudits initialized in the $Z$ basis but does create entanglement when applied to the $X$ basis. The latter is true by definition of the entangling gate $G_E$ in Ref. \cite{quditMBQCstabilizerstates}, which creates maximally entangled states when applied to $\ket{0_X} \ket{0_X}$, so that also $\ket{0_X} \ket{k_X} = \ket{0_X} Z(k) \ket{0_X}$ become maximally entangled due to $Z(k)$ commuting with $G_E$.
To see that no entanglement is created for ancillae in the $Z$ basis, we use that $CZ$ is the only diagonal two-qudit Clifford group generator, so that any diagonal Clifford entangling gate $G_E$ corresponds to $M(N^{-1})CZM(N)$ with $0 \neq N \in \mathbb{F}_d$ (it does not matter whether the multiplication gate $M(N)$, Eq. \eqref{eq:multiplication-gate}, is applied to the control or target) up to local diagonal Clifford gates \cite{quditMBQCstabilizerstates}. Applying this entangling gate to a random $Z$ basis state $\ket{k_Z}$ and $\ket{0_X}$, we then obtain the product state $\ket{k_Z} \otimes Z(Nk) \ket{0_X}$.

Using such resources instead of graph states can then lead to a more efficient decomposition of single-qudit gates into measurement patterns in prime-power dimensions with an even exponent, as discussed in Ref. \cite{quditMBQCstabilizerstates}.

\subsection{Advantages of qudits}
\label{subsec:qudit-advantages}

Due to the larger computational state space provided by finite-dimensional systems, quantum information processing with qudits offers different advantages compared to the binary approach with qubits. For instance, quantum computing can be implemented with reduced circuit depth or resource requirements \cite{efficientalgorithmswithqudits, quditcircuitcompression, ImprovedCircuitDepthTofolli}, allowing certain tasks to be realized more compactly than in qubit-based architectures \cite{QuditComputing}. Furthermore, utilizing qudits has been found to improve the performance of quantum simulations \cite{LatticeGaugeSimulations, quditsimulation_fermionicsystems}, fault-tolerant quantum computing \cite{EnhancedFaultTolerantComputing, FaultTolerantQuditSurfaceCode, QuditsIncreasedStability}, quantum communication \cite{QKDMutuallyUnbiasedBases, QuditQKD_moreSecurity, QuditQKD}, entanglement purification \cite{EntanglementDistillationQudits, EntanglementPurificationQudits}, and quantum metrology \cite{QuantumMetrolgyTransmon, ImprovedSensing}. These advantages are expected to translate to the measurement-based quantum computing framework with qudits, which we have analyzed in Ref. \cite{quditMBQCstabilizerstates}, optimizing measurement patterns for gate implementation.

More specifically, in the context of blind quantum computing, qudits can exhibit a higher robustness with increasing dimension in certain noise models, in particular, for entangled states \cite{QuditComputing}, such as cluster states. In addition, qudits provide a larger hiding space per measurement and per quantum system with increasing dimension.

Since each measurement has $d$ possible outcomes, a malicious server deviating from the prescribed computational protocol can guess the correct outcomes only with probability $(1/d)^n$ for $n$ independently placed trap qudits, assuming outcome indistinguishability guaranteed by blindness. Consequently, higher local dimensions allow the same security threshold to be achieved with fewer trap qudits, reducing the overall resource overhead. The enlarged local measurement space can also allow for more flexible trap and verification constructions.

Overall, the ability to encode more information per quantum system can reduce both classical communication and quantum overhead in client-server architectures. In addition, many experimental platforms naturally support high-dimensional quantum systems \cite{advancesquditentanglement}, rendering qudit-based blind quantum computing particularly appealing from a practical perspective.

\section{Conclusion and outlook}
\label{sec:conclusion}

Blind quantum computing with qudits paves the way for secure delegated quantum information processing using multi-level systems, which are inherent to many experimental platforms.
We have demonstrated how blind quantum computing generalizes from qubits to qudits, how single-qudit gates are kept private by the client introducing random rotations, as well as the overheads of different resource state architectures associated with various hiding strategies for entangling gates. In particular, we have considered brickwork state structures, the open-ended and decorated cluster states, and resource state variants beyond.

To hide single-qudit gates in prime-power dimensions, we have introduced approximately universal gate sets that allow the client to sample and to communicate angles from a finite-sized set. Allowing for continuous-parameter diagonal unitaries, exactly universal gate sets are instead available for qudits in arbitrary dimensions.

To maintain the privacy of entangling gate applications, we have analyzed several resource state architectures. The elementary unit of the brickwork state generalized to qudits is of a fixed size of ten qudits. However, in our qudit variant of the brickwork state, one of the controlled-phase gates is replaced with its inverse. Alternatively, all entangling gates can remain identical to controlled-phase gates at the cost of increasing the elementary unit size with the dimension of the qudit.
Instead, for the open-ended cluster state, the depth of each elementary unit for implementing measurement-based gates increases linearly with the number of qudits, as for qubit blind quantum computing.

Furthermore, we have generalized the decorated cluster state to qudits. Here, a so-called hair of two ancillary qudits can simulate the effect of all necessary measurements on the resource qudit via physical measurements that can be hidden. This hiding strategy also extends to general qudit graph state resources.

Finally, we have discussed graph hiding, where ancillary qudits are initialized in one of two bases that are indistinguishable to the server, such that the controlled-phase gate either does or does not have an effect. This method likewise applies to qudit resources, characterized by diagonal entangling two-qudit Clifford unitaries beyond controlled-phase gates. Using these resources can then lead to increased computational efficiency, facilitating the decomposition of arbitrary single-qudit gates.

In future work, one could compare the security, overhead, and resource cost of the qudit-based scheme with the qubit approach, investigate multi-client settings, or the feasibility of blind quantum computing with reduced quantum capabilities on the client side. Moreover, it would be interesting to embed blind quantum computing into fault-tolerant architectures and analyze their noise resilience.

\begin{acknowledgments}
We acknowledge support from the Austrian Research Promotion Agency (FFG) under Contract Number FO999914030‌ (Next Generation EU). In addition, this research was funded in whole or in part by the Austrian Science Fund (FWF) Grants No. 10.55776/P36009, No. 10.55776/P36010, and No. 10.55776/COE1. Finanziert von der Europäischen Union.
\end{acknowledgments}

\interlinepenalty=10000

\bibliographystyle{apsrev4-2}
\bibliography{quditBQC}

\interlinepenalty=0

\clearpage

\begin{appendix}

\section{Integer-ring qudits}
\label{app:integer-ring-qudits}

In any finite dimension $d$, the qudit basis states can be identified with elements of the integer ring
\begin{equation*}
    \mathbb{Z}_d = \{0,\hdots ,d-1 \},
\end{equation*}
in which addition and multiplication are performed modulo $d$ \cite{QuditComputing,LinearizedStabilizerFormalism, QuditsArbitraryDimension}.

The generalized Pauli operators are then defined via
\begin{equation*}
    Z_d \ket{j} = (\omega_{d})^j \ket{j}, \quad X_d \ket{j} = \ket{j+1},
\end{equation*}
where $\omega_{d} = e^{2\pi i/d}$ is the $d$-th root of unity, and they satisfy the commutation relation \cite{LinearizedStabilizerFormalism}
\begin{equation*}
    X_d^b Z_d^a = \omega_d^{-ab}  Z_d^a X_d^b.
\end{equation*}

The integer-ring single-qudit Clifford group that maps Pauli operators onto Pauli operators under conjugation is generated by $Z_d$ and the gates \cite{Farinholt_2014}
\begin{equation*}
    H_d = \frac{1}{\sqrt{d}} \sum_{j,k = 0}^{d-1} \omega_{d}^{kj} \ket{j} \bra{k}, \quad S_d = \sum_{j =0 }^{d-1} \tau_d^{j^2} \ket{j} \bra{j}
\end{equation*}
with $\tau_d = (-1)^d e^{\frac{\pi i}{d}}$, the Hadamard gate $H_d$, and the phase gate $S_d$.
Extending the single-qudit Clifford group with the controlled-$Z_d$ gate
\begin{equation*}
    CZ_d = \sum_{k,j=0}^{d-1} \omega_d^{kj} \ket{k} \ket{j} \bra{k} \bra{j},
\end{equation*}
yields the $n$-qudit integer-ring Clifford group for arbitrary finite dimensions $d$ \cite{Farinholt_2014,QuditsArbitraryDimension,BenchmarkingUniversalQuditGateset}.

As for finite-field qudits, the generalized $X_d$ and $Z_d$ gates are no longer self-adjoint, but their respective eigenstates still form an orthonormal basis of the qudit Hilbert space.
More specifically, the eigenvectors of $X_d$ correspond to
\begin{equation*}
    \ket{k_X} = H_d \ket{k_Z} = H_d X_d^k \ket{0_Z} = Z_d^k \ket{0_X},
\end{equation*}
whereas the generalized $Y_d$ operator has eigenstates $\ket{k_Y} = S_d \ket{k_X} = S_d H_d \ket{k_Z}$.

\section{Clifford conjugation relations}
\label{app:clifford-conjugation}

Considering qudits of prime-power dimension, $d=p^m$ with $p$ prime and $m$ being a positive integer, we provide in the following some conjugation relations, also derived in Ref. \cite{quditMBQCstabilizerstates}.

The multiplication gate $M(\lambda)$, Eq. \eqref{eq:multiplication-gate}, modifies the Pauli gates via
\begin{equation*}
    Z(z) \xmapsto{M(\lambda)} Z(\lambda^{-1}z), \quad X(x) \xmapsto{M(\lambda)} X(\lambda x),
\end{equation*}
where $x,z \in \mathbb{F}_{d}$.

The Hadamard gate conjugates the Pauli operators according to
\begin{equation*}
    \begin{aligned}
        & H Z(z) H^\dagger =  X(-z), \quad H X(x) H^\dagger =  Z(x).
    \end{aligned}
\end{equation*}

The phase gate commutes with $Z(z)$, whereas in odd prime-power dimensions with $p \neq 2$, one obtains
\begin{equation*}
\begin{aligned}
    & S(\lambda) X(x) S(\lambda)^{-1} = \chi(2^{-1} \lambda x^2 ) X(x) Z(\lambda x),
\end{aligned}
\end{equation*}
while for $p=2$
\begin{equation*}
\begin{aligned}
    & S X(x) S^{-1} = \chi_4(x^2) X(x) Z(x).
\end{aligned}
\end{equation*}

The controlled-$Z$ gate $CZ$ commutes with $Z(z)$ and transforms
\begin{equation*}
\begin{aligned}
    CZ (X(x) \otimes I_d) CZ^\dagger = X(x) \otimes Z(x).
\end{aligned}
\end{equation*}

For the integer-ring qudits in Appendix \ref{app:integer-ring-qudits}, similar conjugation relations hold \cite{LinearizedStabilizerFormalism}.

\section{Non-Clifford diagonal gates}
\label{app:nonclifford-diagonal}

\subsection{Even prime-powers}

To lift the qubit $T$ gate to even prime-power dimensions, we use that $\mathbb{Z}_8 / \langle 2 \rangle \cong \mathbb{Z}_2 $ (since $\langle 2 \rangle = \{2,4,6 \}$ is a maximal ideal in $\mathbb{Z}_8$; every odd number would generate the whole ring, and of the even elements, the integer two generates the largest ideal).

We observe for invertible $0 \neq x \in \mathbb{F}_d$, the conjugation of $X(x)$ by $T_{2^m}^F$ in Eq. \eqref{eq:even-power-T},
\begin{equation*}
\begin{aligned}
    & T_{2^m}^F X(x) (T_{2^m}^F)^\dagger
    = \sum_{u \in \mathbb{F}_d} \chi_8((u+x)^4) \ket{u+x} \bra{u} \chi_8(- u^4)
     \\ & = \sum_{u \in \mathbb{F}_d} \chi_8(x^4 + 4(u^3 x+x u^3)+6x^2 u^2) \ket{u+x} \bra{ u}
    \\ &= \chi_8(x^4) \sum_{u \in \mathbb{F}_d} \chi(u^3 x+x u^3) \ket{u+x} \bra{ u} \chi_8(6x^2 u^2)
    \\ &= \chi_8(x^4) \sum_{u \in \mathbb{F}_d} \ket{u+x} \bra{ u} \chi_4(3x^2 u^2)
    \\ & = \chi_8(x^4) \sum_{u \in \mathbb{F}_d} \ket{u+x} \bra{ u} \chi_4(x^2 u^2) \chi(x u)
    \\ & = \chi_8(x^4) X(x) M(x^{-1}) S M(x) Z(x)
\end{aligned}
\end{equation*}
where we used that $\chi_8(2t) = \chi_4(t) $, $\chi_8(4t) = \chi(t)$, $\chi(u^3 x+x u^3) = \chi(2ux)= \chi(0)= 1$ and $\chi_4(3x^2 u^2) = \chi_4(2x^2 u^2+x^2 u^2) = \chi_4(x^2 u^2) \chi(x^2 u^2) = \chi_4(x^2 u^2) \chi(x u) $.

The result is not a Pauli operator, so we have found a non-Clifford diagonal gate, which reproduces the qubit $T$ gate for $m=1$.

\subsection{Prime-powers with prime three}

To lift the qutrit $T_3$ gate to prime-power dimensions with prime three, we use that $\mathbb{Z}_9 / \langle 3 \rangle \cong \mathbb{Z}_3 $ (since $\langle 3 \rangle = \{3,6 \}$ is a maximal ideal in $\mathbb{Z}_9$).

Conjugating $X(x)$, $0 \neq x \in \mathbb{F}_d$, by the proposed $T_{3^m}^F$ gate in Eq. \eqref{eq:three-power-T}, we obtain
\begin{equation*}
\begin{aligned}
    & T_{3^m}^F X(x) (T_{3^m}^F)^\dagger = \sum_{u \in \mathbb{F}_d} \chi_9((u+x)^3) \ket{u+x} \bra{u} \chi_9(- u^3)
    \\ & = \sum_{u \in \mathbb{F}_d} \chi_9(x^3 + 3 u^2 x + 3 x^2 u) \ket{u+x} \bra{ u}
    \\ & = \chi_9(x^3) \sum_{u \in \mathbb{F}_d} \chi( u^2 x + x^2 u) \ket{u+x} \bra{ u}
    \\ & = \chi_9(x^3) \sum_{u \in \mathbb{F}_d} \ket{u+x} \bra{ u} S(2x) Z(x^2)
    \\ & = \chi_9(x^3) X(x) S(2x) Z(x^2).
\end{aligned}
\end{equation*}
Here, we have used $\chi_9(3x) = \chi(x)$ and that $x^3 = x$ for all $x \in \mathbb{F}_{3^m}$ due to the characteristic $p$ being three.

The result is not a Pauli operator, so that we have a diagonal non-Clifford gate, which coincides with $T_3$ from Eq. \eqref{eq:qutrit-Tgate} for $m=1$.

\subsection{Other prime-powers}

For prime-power dimensions $d= p^m$ with $p \notin \{2,3 \}$, we take the suggested $T_d^F$ gate from Eq. \eqref{eq:T-other-prime-powers}, such that the conjugation $T_d^F X(x) (T_d^F)^\dagger$ returns
\begin{equation*}
\begin{aligned}
    & \sum_{u \in \mathbb{F}_d} \chi(6^{-1} (u+x)^3) \ket{u+x} \bra{u} \chi(-6^{-1} u^3)
    \\ & = \sum_{u \in \mathbb{F}_d} \chi(6^{-1} (x^3 + 3 u^2 x + 3 x^2 u)) \ket{u+x} \bra{ u}
    \\ & = \chi(6^{-1} x^3) \sum_{u \in \mathbb{F}_d} \chi(2^{-1} u^2 x + 2^{-1} x^2 u) \ket{u+x} \bra{ u}
    \\ & = \chi(6^{-1} x^3) \sum_{u \in \mathbb{F}_d} \ket{u+x} \bra{ u} S(x) Z(2^{-1} x^2)
    \\ & = \chi(6^{-1} x^3) X(x) S(x) Z(2^{-1} x^2).
\end{aligned}
\end{equation*}
Since the result is not a Pauli operator, we have found a diagonal non-Clifford gate $T_d^F$.

\section{Measurement patterns on the qudit brickwork state}

In the following, we show that the qubit measurement patterns to implement a Hadamard gate, Fig. \ref{fig:brickwork-patterns} $(a)$, and a controlled-$X$ gate, Fig. \ref{fig:brickwork-patterns} $(b)$, generalize to the qudit brickwork state elementary unit, Fig. \ref{fig:brickwork-resource}. For notational simplicity, we write $S$ not only for the even prime-power dimensional phase gate but also for $S(1)$ in the odd-dimensional case.

\subsection{Hadamard gate implementation}
\label{app:brickwork-Hadamard}

Using the measurement pattern, displayed in Fig. \ref{fig:brickwork-patterns} $(a)$ and generalized to qudits, one can implement the Hadamard gate on either of the two logical qudits (where the first entry in the tensor product represents the upper logical qudit and the second the lower) due to
\begin{equation*}
\begin{aligned}
    & CZ^{-1} (H^2 S \otimes H^2) CZ (H S H S \otimes H^2) 
    \\ & = CZ^{-1} (M(-1)  \otimes M(-1)) CZ ( SH S H S \otimes M(-1)) 
    \\ & = CZ^{-1} (M(-1)  \otimes M(-1)) CZ ( H^{-1} \otimes M(-1)) 
    \\ & = CZ^{-1} (M(-1)  \otimes M(-1)) CZ (H^2 \otimes M(-1)) (H \otimes I)
    \\ & =  CZ^{-1} CZ (H \otimes I) = H \otimes I.
\end{aligned}
\end{equation*}
Here, we used that $H^2 = M(-1)$, $H^{-1} = H^3$ and the identity $H^{-1} = SH S H S$, which follows from the multiplication gate decomposition of $M(1) = I_d$ in Eq. \eqref{eq:multiplication-gate-decomposition}.

\subsection{Controlled-$X$ gate implementation}
\label{app:brickwork-CX}

To implement a controlled-$X$ gate on the qudit version of the brickwork state, we use the measurement pattern, displayed in Fig. \ref{fig:brickwork-patterns} $(b)$ and generalized to qudits, which realizes the gate sequence
\begin{equation*}
\begin{aligned}
    & CZ^{-1} ( H^2 S \otimes H S^{-1} H ) CZ  (H^2 \otimes H S H ) 
    \\ & = CZ^{-1} ( M(-1) S \otimes H S^{-1} H ) CZ  (M(-1) \otimes H S H )
    \\ & = CZ^{-1} ( M(-1) S \otimes H S^{-1} ) CX^{-1}_{1 \mapsto 2}  (M(-1) \otimes M(-1) S H ) 
    \\ & = CZ^{-1} ( S \otimes H S^{-1} ) CX_{1 \mapsto 2}  (I \otimes S M(-1)  H ) 
    \\ & = CZ^{-1} ( I \otimes H ) CX_{1 \mapsto 2} CZ^{-1}  (I  \otimes H^{-1} )
    \\ & = CZ^{-1} CZ CX_{1 \mapsto 2} = CX_{1 \mapsto 2}.
\end{aligned}
\end{equation*}

Note that $S$ and $M(-1)$ commute (since the phase in $S$ is specified by the square of the basis state, so $M(-1) S M(-1) = S$) and $M(-1) H = H^3 = H^{-1}$.

Furthermore, we use in the above calculation that
\begin{equation*}
\begin{aligned}
    & (S \otimes S^{-1}) CX_{1 \mapsto 2}
    = (S \otimes S^{-1}) \sum_{x} \ket{x} \bra{x} \otimes X(x)
    \\ & = \left(\sum_{x} \ket{x} \bra{x} \otimes \chi(-2^{-1} x^2) X(x)Z(-x) \right) (S \otimes S^{-1})
    \\ & = \left(\sum_{x} \ket{x} \bra{x} \otimes X(x)Z(-x) \right) (I \otimes S^{-1})
    \\ & = CX_{1 \mapsto 2} CZ^{-1} (I \otimes S^{-1}),
\end{aligned}
\end{equation*}
so that $(S \otimes S^{-1}) CX_{1 \mapsto 2} (I \otimes S) = CX_{1 \mapsto 2} CZ^{-1}$.

\section{Qudit mirror operator on the open-ended cluster state}
\label{app:qudit-mirror}

We want to prove that for an even number $n$ of logical qudits, we have a global mirror operator after $n+1$ applications of the qudit variant of $C_n$ from Eq. \eqref{eq:layer-operator} while, for odd $n$, we have a global mirror plus multiplication with minus one, $M(-1) = H^2$, on every qudit. This generalizes the qubit mirror from Ref. \cite{clusteruniversality-XYmeasurements}.

We show this by demonstrating that the Pauli gates $Z$ and $X$ acting on any of the input qudits are reflected when propagated by $C_n$ towards the output if $n$ is even and reflected and conjugated to $Z^{-1}$ and $X^{-1}$, respectively, if $n$ is odd. Here, for prime-power dimensions, $Z^{-1}$ and $X^{-1}$ are $Z(-1)$ and $X(-1)$ instead, but we use the prime-dimensional notation for conciseness in the following.

To understand how the effective quantum circuit conjugates the input Paulis, we first consider $Z_1$ acting on the first qudit of the input. Due to repeated conjugation with $C_n$, corresponding to $X$ measurements in one column of the open-ended cluster state, the Pauli gets conjugated according to
\begin{equation*}
    Z_1 \xmapsto{C_n} X_1^{-1} \xmapsto{C_n} Z_1^{-1} X_2 \xmapsto{C_n} Z_2 X_3^{-1} \xmapsto{C_n} Z_3^{-1} X_4 \hdots,
\end{equation*}
which continues until the string reaches qudit $n$ after $n$ steps. If $n$ is even, after $n$ steps, the stabilizer is given by $Z_{n-1}^{-1} X_{n}$ while if $n$ is odd, it is $Z_{n-1} X_{n}^{-1}$. The next conjugation then changes this to 
\begin{equation*}
    Z_{n-1}^{-1} X_{n} \xmapsto{C_n} Z_n
\end{equation*}
for the even case and for $n$ odd to 
\begin{equation*}
    Z_{n-1} X_{n}^{-1} \xmapsto{C_n} Z_n^{-1}.
\end{equation*}

If we now consider the second qudit of the input,
\begin{equation*}
\begin{aligned}
    & Z_2 \xmapsto{C_n} X_2^{-1} \xmapsto{C_n} X_1 Z_2^{-1} X_3 \xmapsto{C_n} Z_1 X_2^{-1} Z_3 X_4^{-1}
    \\ & \xmapsto{C_n} Z_2^{-1} X_3 Z_4^{-1} X_5 \xmapsto{C_n} Z_3 X_4^{-1} Z_5 X_6^{-1} \hdots,
\end{aligned}
\end{equation*}
so that after three applications of $C_n$, we have a Pauli string extending to the first qudit that subsequently begins to shift.

If $n$ is even, after $n-1$ applications ($n-4$ to shift the Pauli string and three prior) we have the stabilizer $Z_{n-3} X_{n-2}^{-1} Z_{n-1} X_n^{-1}$ while for odd $n$, it is $Z_{n-3}^{-1} X_{n-2} Z_{n-1}^{-1} X_n$. Then, two further columns being measured for $n$ even results in
\begin{equation*}
\begin{aligned}
    & Z_{n-3} X_{n-2}^{-1} Z_{n-1} X_n^{-1} \xmapsto{C_n} Z_{n-2}^{-1} X_{n-1} Z_n^{-1},
    \xmapsto{C_n} Z_{n-1}
\end{aligned}
\end{equation*}
so that, in total $Z_2 \mapsto Z_{n-1}$ for $n$ even and $Z_2 \mapsto Z_{n-1}^{-1}$ for $n$ odd.

Considering the third qudit of the input, $X$ measurements in the first five columns result in
\begin{equation*}
\begin{aligned}
    & Z_3 \xmapsto{C_n} X_3^{-1} \xmapsto{C_n} X_2 Z_3^{-1} X_4 \xmapsto{C_n} X_1^{-1} Z_2 X_3^{-1} Z_4 X_5^{-1}
    \\ & \xmapsto{C_n} Z_1^{-1} X_2 Z_3^{-1} X_4 Z_5^{-1} X_6 \xmapsto{C_n}  Z_2 X_3^{-1} Z_4 X_5^{-1} Z_6 X_7^{-1},
\end{aligned}
\end{equation*}
so that we again have a Pauli string first spreading until it reaches the first qudit and subsequently shifting.

More generally, for the $m$-th input qudit in the first column and row $m < \frac{n+1}{2}$, it takes $m+1$ steps until either the Pauli string $Z_1 X_2^{-1} \hdots Z_{2m-1} X_{2m}^{-1}$ for $m$ even or $Z_1^{-1} X_2 \hdots Z_{2m-1}^{-1} X_{2m}$ for $m$ odd has appeared.
Then, it takes another $n - 2m$ steps to shift the Pauli string to act on qudit $n$. For $n$ and $m$ even, we then have after $n-m+1$ steps the Pauli string
\begin{equation*}
    Z_{n-2m+1} X_{n-2m+2}^{-1} \hdots Z_{n-1} X_n^{-1}.
\end{equation*}
The next two conjugations with $C_n$ change this to
\begin{equation*}
\begin{aligned}
    & \xmapsto{C_n} Z_{n-2m+2}^{-1} X_{n-2m+3} \hdots X_{n-1} Z_n^{-1}
    \\ & \xmapsto{C_n} Z_{n-2m+3} X_{n-2m+4}^{-1} \hdots X_{n-2}^{-1} Z_{n-1},
\end{aligned}
\end{equation*}
so that the Pauli string begins to shrink until it becomes $Z_{n-m+1}$ after $m$ steps.
This is essentially the reverse procedure of the Pauli string expanding during the first $m+1$ steps.
For $n$ even and $m$ odd, the same follows since the shifted Pauli after $n-m+1$ steps has opposite powers, but $m$ odd during the shrinkage phase reverses the powers back.
Instead, for $n$ odd the Pauli becomes $Z_{n-m+1}^{-1}$, so that in addition to the global mirror a multiplication with minus one has happened, $Z_{n-m+1} \xmapsto{M(-1)} Z_{n-m+1}^{-1}$.

For $X_m$ on the $m$-th input qudit, we require one less steps, $m$ steps, in the Pauli string expansion phase and one more step after the shrinkage phase, so that with the same argument as previously, the Pauli transforms into $X_{n-m+1}$ for $n$ even and $X_{n-m+1}^{-1}$ for $n$ odd.

Thus, we have identified that $n+1$ applications of $C_n$ act as a global mirror on the qudit input for $n$ even and, in addition, apply $M(-1)^{\otimes n}$ for $n$ odd.

\section{Entangling gate between neighboring logical qudits of the open-ended cluster state}
\label{app:entangling-gate-clusterstate}

We consider an open-ended cluster state of lattice size $n \cross (n+2)$, Figs. \ref{fig:hiding-entangling-gate} $(a)$ and \ref{fig:openended-clusterstate}, where $C_n$, Eq. \eqref{eq:layer-operator} with the gates generalized to qudit operators, is applied for each column being measured in the $X$ basis.

Studying how a diagonal gate $D_{\Vec{\alpha}}$ in the first row and a column $m$ with $1 < m < n+1$ propagates under conjugation by $C_n$ which is correspondingly applied $n+2 - m$ times,
\begin{equation*}
\begin{aligned}
    & C_n^{\otimes n+1 - m} \prod_{j=1}^n H_j D_{\Vec{\alpha}} \prod_{j=1}^{n-1} CZ_{j,j+1} C_n^{\otimes m-1}
    \\ & = C_n^{\otimes n+2 - m} D_{\Vec{\alpha}} C_n^{\otimes m-1}
    \\ & =  C_n^{\otimes n+2 - m} D_{\Vec{\alpha}} (C_n^{\otimes n+2 - m})^\dagger C_n^{\otimes n+1},
\end{aligned}
\end{equation*}
we see in the following that an entangling gate on the qudits $n+1-m$ and $n+2 - m$ of the output is realized.

Any diagonal unitary gate can be written via
\begin{equation*}
    D_{\Vec{\alpha}} = e^{i \sum_k \alpha_k \ket{k_Z} \bra{k_Z}}.
\end{equation*}
After the first conjugation with $C_n$, due to commutation of $D_{\Vec{\alpha}}$ with $CZ$, we map the diagonal gate onto
\begin{equation*}
    H D_{\Vec{\alpha}} H^\dagger = e^{i \sum_k \alpha_k \ket{k_X} \bra{k_X}}.
\end{equation*}
After the second conjugation with $C_n$, we have
\begin{equation*}
\begin{aligned}
    & (H \otimes H) CZ \left( e^{i \sum_k \alpha_k \ket{k_X} \bra{k_X} \otimes I_d} \right) CZ^\dagger (H^{\dagger} \otimes H^{\dagger})
    \\ & = (H \otimes H) \left( e^{i \sum_{k,j} \alpha_k \ket{(k+j)_X} \bra{(k+j)_X} \otimes \ket{j_Z} \bra{j_Z}} \right) (H^{\dagger} \otimes H^{\dagger})
    \\ & = \left( e^{i \sum_{k,j} \alpha_k \ket{-(k+j)_Z} \bra{-(k+j)_Z} \otimes \ket{j_X} \bra{j_X}} \right),
\end{aligned}
\end{equation*}
where we use that $H^2 = M(-1)$, so that $H \ket{k_X} = \ket{-k_Z}$.
Repeating the conjugation with $C_n$, we obtain
\begin{equation*}
\begin{aligned}
    & e^{i \sum_{k,j,l} \alpha_k \ket{-(k+j)_Z} \bra{-(k+j)_Z} \otimes \ket{j_X} \bra{j_X} \otimes \ket{l_Z} \bra{l_Z}}
    \\ & \xmapsto{CZ^{\otimes 2}}  e^{i \sum_{k,j,l} \alpha_k \ket{-(k+j)_Z} \bra{-(k+j)_Z} \otimes \ket{(l-k)_X} \bra{(l-k)_X} \otimes \ket{l_Z} \bra{l_Z}}
    \\ & \xmapsto{H^{\otimes 3}}  e^{i \sum_{k,j,l} \alpha_k \ket{-(k+j)_X} \bra{-(k+j)_X} \otimes \ket{(k-l)_Z} \bra{(k-l)_Z} \otimes \ket{l_X} \bra{l_X}}
    \\ & =  e^{i \sum_{k,l,j} \alpha_k Z^{-k} \ket{-j_X} \bra{-j_X} Z^k \otimes \ket{(k-l)_Z} \bra{(k-l)_Z} \otimes \ket{l_X} \bra{l_X}}
    \\ & =  e^{i \sum_{k,l} \alpha_k I_d \otimes \ket{(k-l)_Z} \bra{(k-l)_Z} \otimes \ket{l_X} \bra{l_X}}.
\end{aligned}
\end{equation*}
Hence, now no operator acts on the first qudit, but we instead obtain an entangling gate on the second and third qudits.

Another application of $C_n$ yields
\begin{equation*}
\begin{aligned}
    & e^{i \sum_{k,l,j} \alpha_k \ket{(k-l)_Z} \bra{(k-l)_Z} \otimes \ket{l_X} \bra{l_X} \otimes \ket{j_Z} \bra{j_Z}}
    \\ & \xmapsto{CZ^{\otimes 2}} e^{i \sum_{k,l,j} \alpha_k \ket{(k-l)_Z} \bra{(k-l)_Z} \otimes \ket{(k+j)_X} \bra{(k+j)_X} \otimes \ket{j_Z} \bra{j_Z}}
    \\ & = e^{i \sum_{k,j} \alpha_k I_d \otimes \ket{(k+j)_X} \bra{(k+j)_X} \otimes \ket{j_Z} \bra{j_Z}}
    \\ & \xmapsto{H^{\otimes 3}} e^{i \sum_{k,j} \alpha_k I_d \otimes \ket{-(k+j)_Z} \bra{(k+j)_Z} \otimes \ket{j_X} \bra{j_X}},
\end{aligned}
\end{equation*}
so that the entangling interaction has moved to the next pair of qudits.

For an even number of applications of $C_n$, so $n+2 - m$ even, we then have the entangling gate
\begin{equation*}
\begin{aligned}
    & e^{i \sum_{k,j} \alpha_k \ket{-(k+j)_Z} \bra{-(k+j)_Z} \otimes \ket{j_X} \bra{j_X}},
\end{aligned}
\end{equation*}
whereas for an odd number of applications of $C_n$, it is 
\begin{equation*}
\begin{aligned}
    & e^{i \sum_{k,j} \alpha_k \ket{(k-j)_Z} \bra{(k-j)_Z} \otimes \ket{j_X} \bra{j_X}}.
\end{aligned}
\end{equation*}

\end{appendix}

\end{document}